# Investigation of GeSn aspect ratio trapping growth up to 8% Sn

*Hryhorii Stanchu [1‡], Quang Minh Thai [2‡], Fernando M. de Oliveira [1], Mourad Benamara [1], Stephen Margiotta [3], Matthew Cook [3], Xiaoxin Wang [4], Jifeng Liu [4], Perry C. Grant [5], Baohua Li [5], Wei Du [1,2,6], Gregory Salamo [1,6,7] and Shui-Qing Yu [1,2,6*]*

[1] Arkansas Materials Institute, University of Arkansas, Fayetteville, Arkansas 72701, USA

[2] Department of Electrical Engineering and Computer Science, University of Arkansas, Fayetteville, Arkansas 72701, USA

[3] Lincoln Laboratory, Massachusetts Institute of Technology, 244 Wood Street, Lexington, Massachusetts 02421, USA

[4] Thayer School of Engineering, Dartmouth College, Hanover, New Hampshire 03755, USA

[5] Arktonics, LLC, 1339 South Pinnacle Dr., Fayetteville, Arkansas 72701, USA

[6] Material Science and Engineering Program, University of Arkansas, Fayetteville, Arkansas 72701, USA

[7] Department of Physics, University of Arkansas, Fayetteville, Arkansas 72701, USA

**ABSTRACT**. Aspect ratio trapping (ART) growth of germanium-tin (GeSn) is a promising approach to target important objectives on the quest towards commercialization of complementary metal-oxide-semiconductor (CMOS)-compatible GeSn optoelectronics devices. Its local growth on patterned substrate allows for versatile device integration into photonics integrated circuit or for stand-alone structure like focal plane array imager. Additionally, high aspect ratio from nano-sized window can terminate early threading dislocation propagation on the oxide sidewalls, leaving subsequent growth defect-free and potentially improving the device performance. Knowledge remains missing regarding GeSn ART growth kinetics, morphology and how they evolve from thin film growth, with successful growth itself yet to be demonstrated. In this work, we report GeSn ART growth up to 8% Sn. Two configurations – self-induced Ge core/GeSn shell for Sn content between 6% and 8%, and bulk GeSn ART for Sn content below 1% – are observed. We present a comprehensive study on GeSn ART growth kinetics through different growth rounds and designs, showing a link between pyramid shape of ART island and successful Sn incorporation, as well as the role of growth selectivity and local heating.

Research on germanium-tin (GeSn) – group IV semiconductor alloys which possess tunable infrared band gap, can transform into direct gap material under high Sn content and can be monolithically grown on Silicon (Si) – has shown steady progress over the last 10 years, establishing itself as a promising material for the development of complementary metal-oxide-semiconductor (CMOS)-compatible optoelectronics devices. Several milestones for GeSn devices have been reached: GeSn lasers operation have been demonstrated at room temperature under optical pumping condition [1–3], with progress towards electrically-injected, continuous wave operation at room temperature [4–7]. At the same time, GeSn detectors gradually established its position in extended short-wavelength infrared (SWIR) and low mid-wavelength infrared (MIR) range, reducing the performance gap with those from III-V and II-VI alloys, with demonstration of GeSn infrared imager/ gas detector prototype and more importantly, a better understanding on their intrinsic device physics and performance bottlenecks to formulate efficient optimization strategy [8–14]. With these results, research vision for GeSn now turns towards its technology maturity, aiming for a balance between device performance, production cost and scalability suitable for future commercialization.

Recent works clearly reflected this trend, with research direction focusing on optimizing GeSn growth quality under conditions suitable for foundry mode operation, using isothermal, flow-variable GeSn/SiGeSn heteroepitaxy [15] or pivoting from exotic and costly Ge gas precursor to mainstream, industrial standard germane gas ($GeH_4$) [15,16]. Another direction focused on the integration aspect of GeSn optoelectronics devices: recent work from Stanchu *et al.* demonstrated GeSn selective area growth (SAG) with Sn content up to 8.7% [17]. Such local growth on patterned substrate offered a viable pathway for integration of GeSn lasers/ detectors into CMOS fabrication line, suitable for both co-integration of GeSn lasers/detectors with Si-based waveguides in photonics integrated circuit (PIC) or stand-alone module like focal plane array (FPA) imager [18]. Early results so far revealed a drastic change of growth condition between SAG with small oxide window size (below 10 μm) and normal thin film growth, with higher growth rate making high Sn incorporation in small window size SAG particularly challenging [17]. Adjusting the growth recipe to address these issues, and scaling GeSn SAG to nano-sized window – also known as aspect ratio trapping (ART) growth – is of particular interest: here, narrow oxide window with high aspect ratio (i.e. ratio of window height over window width higher than 1) can prevent early propagation of threading dislocation at the oxide window sidewalls, leaving subsequent growth defect-free. Previously reported for defect-free, monolithic growth of Ge and III-V alloys on Si [18–23], a successful demonstration of GeSn ART growth can provide the answer for both challenges of CMOS integration and crystal quality in this material.

In this work, we reported GeSn ART growth up to 8% Sn. Two configurations for the overgrown islands were observed: one with self-induced Ge core/ GeSn shell with Sn content between 6% and 8% Sn, the other with bulk Sn incorporation with limited Sn content below 1% Sn. By examining different growth designs and recipes, it was shown that either Ge core/GeSn shell growth recipe, or recipe with low growth selectivity (i.e. with nucleation allowed on oxide surface) increased the chance of Sn incorporation, through the observation of Sn droplets at first, followed by successful GeSn growth. In addition, ART island shapes were found to be strongly correlated with Sn incorporation status: we observed an evolution from dome shape (no Sn incorporation) to polygon with multi-facets shape (presence of Sn droplets) to pyramid shape (Sn droplets or successful GeSn growth).

## RESULTS AND DISCUSSION

### ART growth design and samples summary

Ge and GeSn ART growth were conducted on Si wafer with patterned $SiO_2$ window, using a commercial ASM Epsilon 2000 reduced pressure chemical vapor deposition (RPCVD) reactor. $SiO_2$ mask layer thickness was 550 nm, with three different diameters for circular window openings: 180 nm, 280 nm and 1 µm. $GeH_4$ and $SnCl_4$ were selected for Ge and Sn gas precursors, respectively. HCl flow was introduced during the growth and adjusted to control the growth selectivity, i.e. the ability to inhibit Ge, GeSn or Sn nucleation on $SiO_2$ surface. Three different GeSn ART growth designs were tested in this work (**Scheme 1**): GeSn baseline, where direct GeSn growth on Si substrate was targeted; Ge core/GeSn shell, where we targeted an overgrowth of Ge out of the oxide window prior to GeSn layer growth; GeSn with short Ge nucleation, where we targeted very thin Ge buffer growth inside the oxide window before GeSn growth. An additional reference Ge ART sample was grown for comparison purpose.

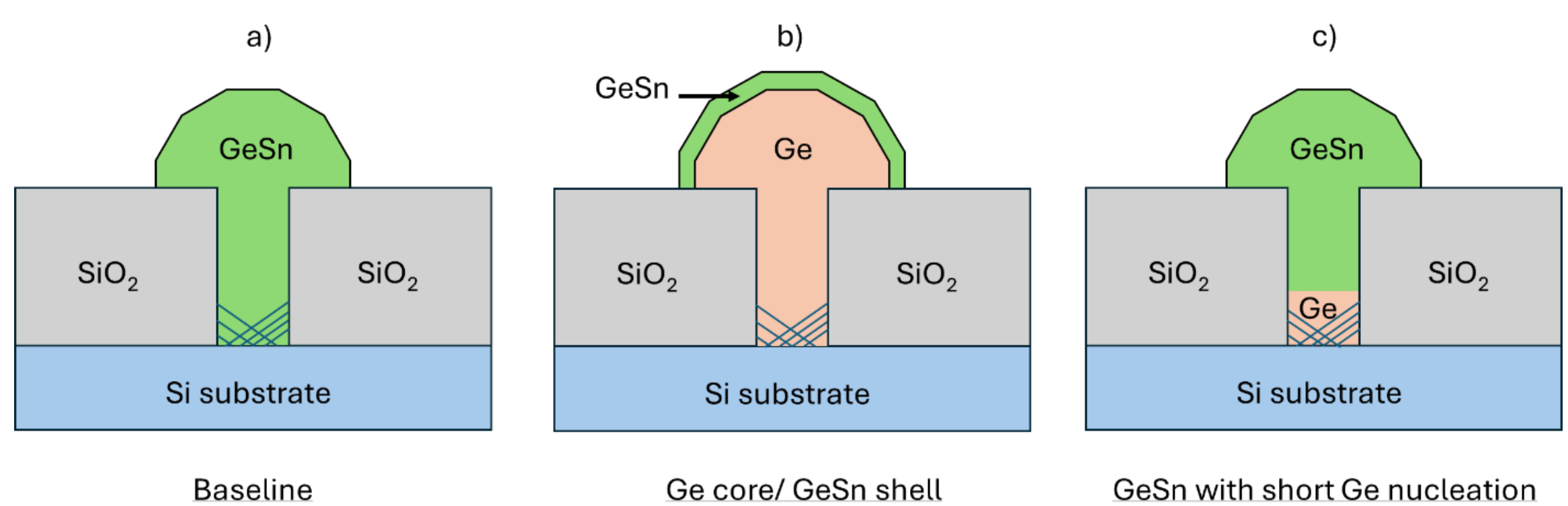

***Scheme 1.*** *Three GeSn ART growth design tested in this work: a) GeSn baseline, b) Ge core/ GeSn shell and c) GeSn growth with short Ge nucleation. Lines inside the oxide window illustrated threading dislocation propagation, which was anticipated to be terminated at the*

*oxide sidewalls. For the actual experimental growth results, please check Table 1 and the rest of the paper.*

Results from five growth rounds were presented in this paper. Round 1 was dedicated to Ge and GeSn baseline ART growth. Rounds 2 and 3 were dedicated to progress in GeSn ART growth through other designs, with the observation of Sn droplets on ART islands. Rounds 4 and 5 were dedicated to successful GeSn ART growth, either in self-induced Ge core/GeSn shell configuration (round 4) or in GeSn bulk configuration (round 5). **Table 1** summarized the results of all Ge and GeSn ART samples presented in this work, with details regarding their ART island shapes, Sn droplets presence at large window (1 µm) and small windows (180 nm, 280 nm), alongside observation of GeSn growth.

***Table 1.*** *Details of all ART samples presented in this work, with their growth design and experimental results.*

| | | | Results | | | | | |
|---|---|---|---|---|---|---|---|---|
| **Round** | **Sample ID** | **Growth design** | **Growth selectivity** | **Shape (1 µm)** | **Sn droplets (1 µm)** | **Shape (180 - 280 nm)** | **Sn droplets (180 - 280 nm)** | **GeSn growth** |
| 1 | 1-A | Baseline (Ge) | High | Dome | N/A | Dome | N/A | N/A |
| | 1-B | Baseline (GeSn) | High | Dome | No | Dome | No | No |
| 2 | 2-A | Ge core/ GeSn shell | High | Octagon, multi-facets | Yes | Octagon, multi-facets | Yes | No |
| | 2-B | GeSn, short Ge nucleation | High | Dome | No | Dome | No | No |
| | 2-C | GeSn, short Ge nucleation | Low | N/A, coalescence | No | Octagon, multi-facets | No | No |
| | 2-D (1) | GeSn, short Ge nucleation | Low | Dome | Yes | Dome | No | No |
| | 2-D (2) | | Low | Circle, multi-facets | Yes | Circle, multi-facets | No | No |
| | 2-D (3) | | Low | Circle, multi-facets | Yes | Circle, multi-facets | No | No |

| | 2-D (4) | | Low | Rounded square, multi-facets | Yes | Pyramid, multi-facets | No | No |
|---|---|---|---|---|---|---|---|---|
| 3 | 3-A | GeSn, short Ge nucleation | Low | Pyramid, {111} facets | Yes | Pyramid, {111} facets | Yes | No |
| 4 | 4-A | GeSn, short Ge nucleation | Low | Dodecagon, multi-facets | Yes | Pyramid, {111} facets | GeSn growth | Yes (6% Sn, shell) * |
| | 4-B | | Low | Octagon, multi-facets | Yes | Pyramid, {111} facets | GeSn growth | Yes (8% Sn, shell) * |
| 5 | 5-A | GeSn, short Ge nucleation | Low | Dodecagon, multi-facets | Yes | Pyramid, {111} facets | GeSn growth | Yes (<1% Sn, bulk) * |

|  |  |
|---|---|
| Red | Unsuccessful ART growth, neither Sn droplets nor GeSn growth was observed |
| Yellow | ART Growth where Sn droplets was observed. GeSn growth was not observed |
| Green | Successful ART growth, with GeSn growth observed on 180 nm/ 280 nm oxide windows |

* Sn content extracted from TEM-EDX data.

**Baseline GeSn ART growth (round 1)**

In the first ART growth round, results from GeSn baseline approach (sample 1-B) was compared to reference Ge ART growth (sample 1-A) (**Figure 1**). SEM images showed high growth selectivity on both samples, without any nucleation on SiO2 surface, confirming the etching efficiency of HCl flux (**Figure 1a**). Smooth, dome shape was observed on Ge ART islands in sample 1-A. On sample 1-B, SEM images showed lower to no lateral overgrowth of ART island, suggesting an increased etching rate in sample 1-B compared to sample 1-A. TEM images (**Figures 1b,c**) further confirmed absence of lateral overgrowth in sample 1-B, and a similar dome shape on both samples. In addition, a noticeable under-etching into the Si substrate was detected in the oxide window: from the TEM image of sample 1-B, one can observe the propagation of some threading dislocation lines, followed by their termination at the oxide walls in this area. No sign of Sn droplets or GeSn growth was detected in sample 1-B.

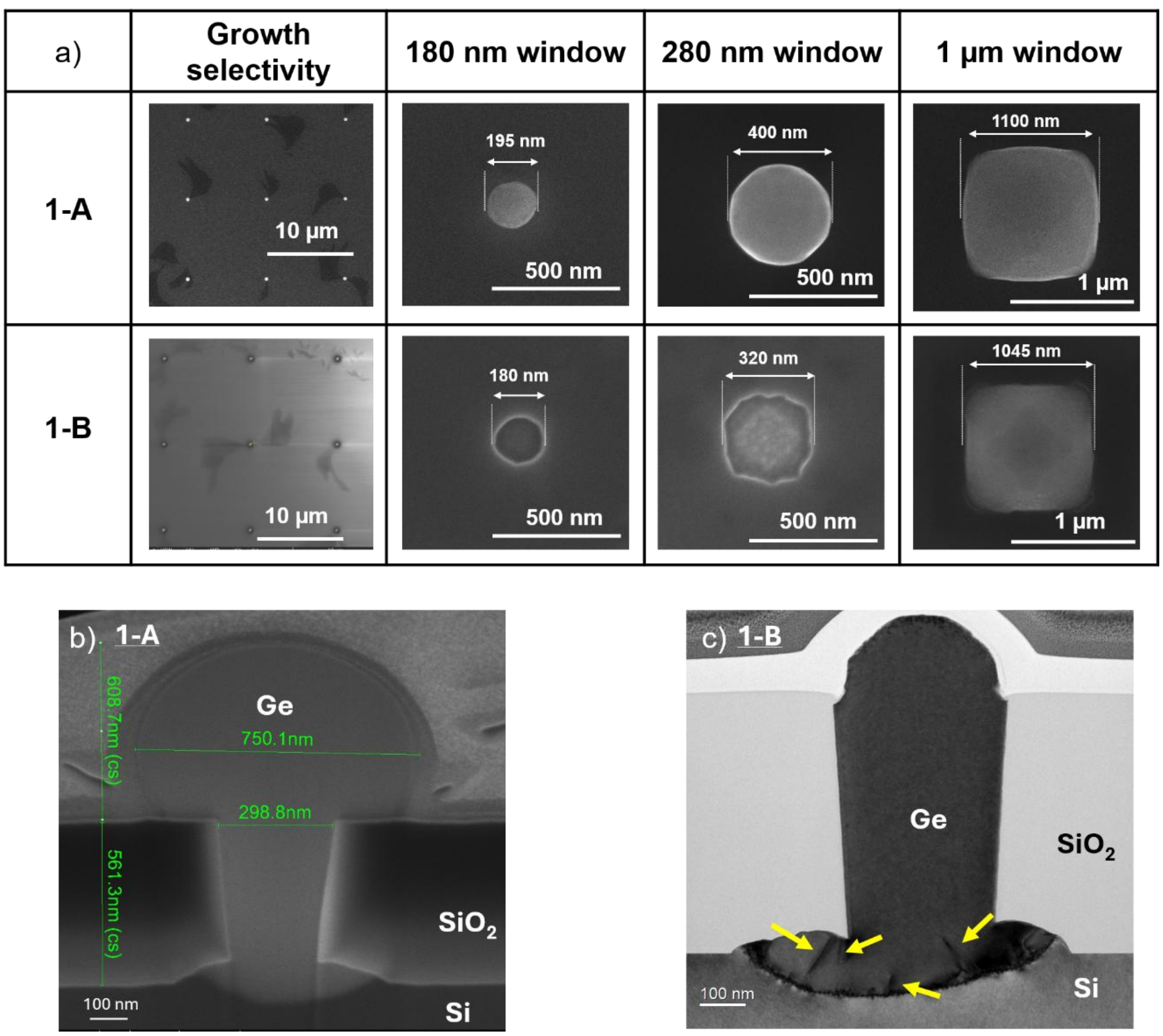

***Figure 1****. a) Zoomed-out SEM image to assess growth selectivity, alongside SEM images of ART islands grown from 180 nm, 280 nm and 1 µm oxide windows of samples 1-A and 1-B. White double-arrows in SEM images showed the size of grown ART island. b), c) TEM images of ART islands grown in 180 nm oxide window from sample 1-A and 1-B, respectively. Yellow arrows pointed to threading dislocation lines located near Si/Ge interface in sample 1-B*

**Progress in GeSn ART growth with observation of Sn droplets (rounds 2-3)**

As GeSn growth was absent from the first round, other ART growth designs were then explored: Ge core/GeSn shell ART, and GeSn ART with short Ge nucleation, both mimicking GeSn thin film growth on Ge buffer layer [24–27]. These designs can provide intermediate, more favorable conditions for Sn incorporation and further insight into GeSn ART growth kinetics. In addition, GeSn ART growth selectivity will be closely monitored and discussed for samples from rounds 2 and 3, as round 1 results suggested that HCl etching rate might be different between Ge and GeSn ART, thus impacting Ge and GeSn nucleation in a different way.

***GeSn ART growth attempt with Ge core/ GeSn shell design (sample 2-A)***

Results for Ge core/GeSn shell growth attempt (sample 2-A) were presented in **Figure 2**. From SEM images (**Figure 2a**), noticeable morphology change was observed compared to previous GeSn baseline growth: multi-facets ART island with a polygon base – octagon in 180 nm/ 280 nm windows and dodecagon in 1 μm window - were observed instead of a dome shape. Sn droplets can be spotted on all ART window sizes from 180 nm to 1 μm, and it was further confirmed by SEM-EDX analysis (**Figures 2b,c**), with most of the droplets located on the side facets {100} and {110} of ART islands. Growth selectivity remained high in sample 2-A, without any nucleation on the oxide surface. These results suggested that under high growth selectivity, a Ge core/GeSn shell growth design was more favorable for Sn nucleation than a baseline GeSn ART growth design.

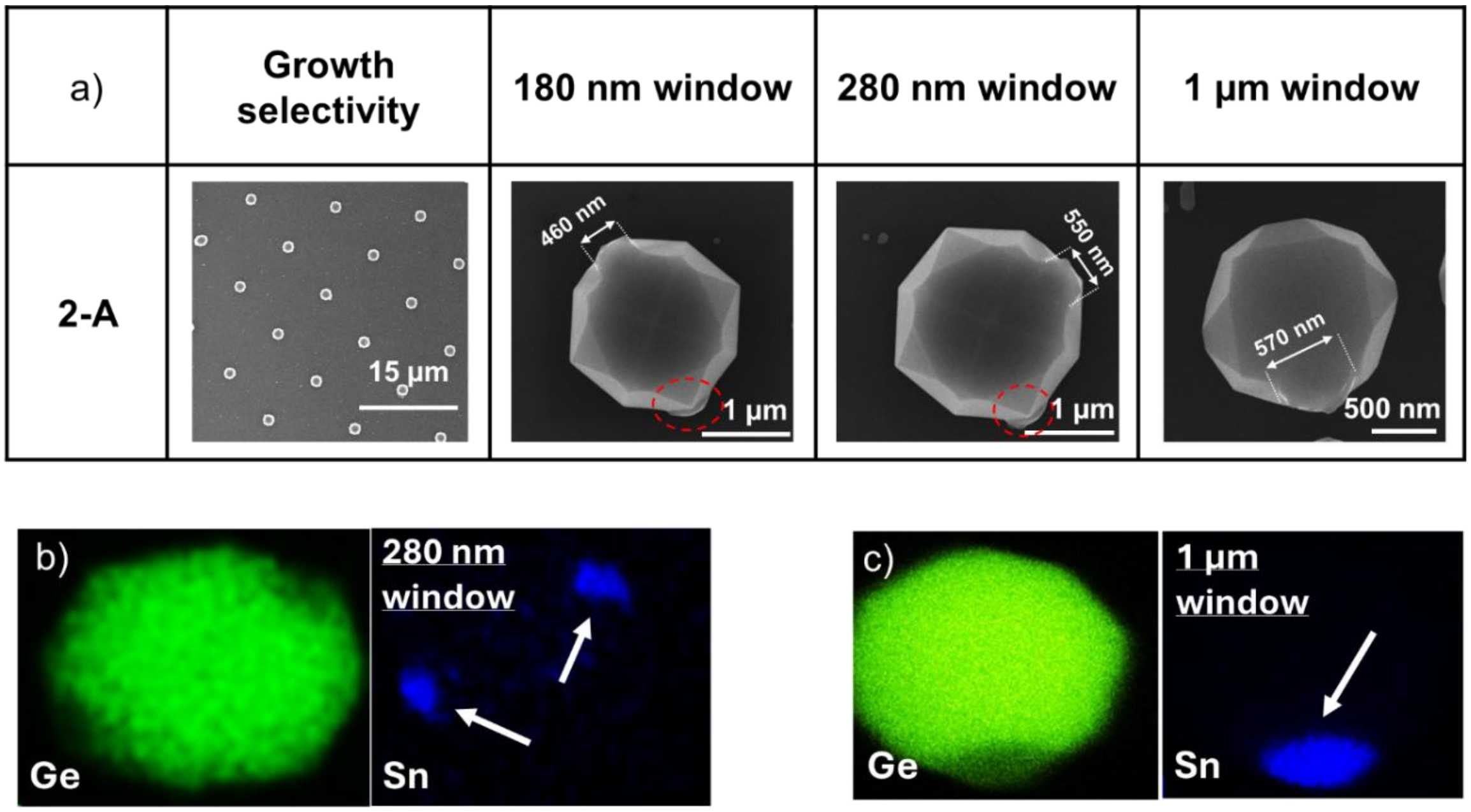

***Figure 2**. a) Zoomed-out SEM image to assess growth selectivity, alongside SEM images of ART islands grown from 180 nm, 280 nm and 1 μm oxide windows of samples 2-A. Red dashed circles helped identifying Sn droplets on ART island. White double-arrows indicated Sn droplet sizes, ranging from 460 nm to 570 nm. b), c) SEM-EDX analysis of ART islands grown on 280 nm and 1 μm windows, respectively. Ge profile was shown in green color, while Sn profile was shown in blue color, with white arrows pointed to Sn droplets. Please note that SEM-EDX analysis were conducted on different ART islands than those shown in a).*

***GeSn ART growth attempt with short Ge nucleation design - Transition to pyramid shape (samples 2-B, 2-C, 2-D, 3-A)***

Several samples were then grown to test GeSn ART growth design with short Ge nucleation inside the oxide window. Results from our earliest growth attempts (2-B, 2-C) were shown in **Figure 3**. Even when no GeSn growth or Sn droplets were observed for either of these samples, a difference in ART island shape and lateral overgrowth was still observed, correlating to a change in growth selectivity: at lower growth selectivity (sample 2-C), ART island shape transformed from dome to octagon, multi-facets (**Figure 3a**), with TEM images showing significant increase in lateral overgrowth (**Figures 3b,c**). Compared to sample 2-A in previous section, the multi-facets feature in sample 2-C was less sharp: under low growth selectivity condition, random Ge nucleation on the oxide surface during lateral overgrowth might compromise the single crystallinity of the ART island and thus can prevent Sn incorporation in this case.

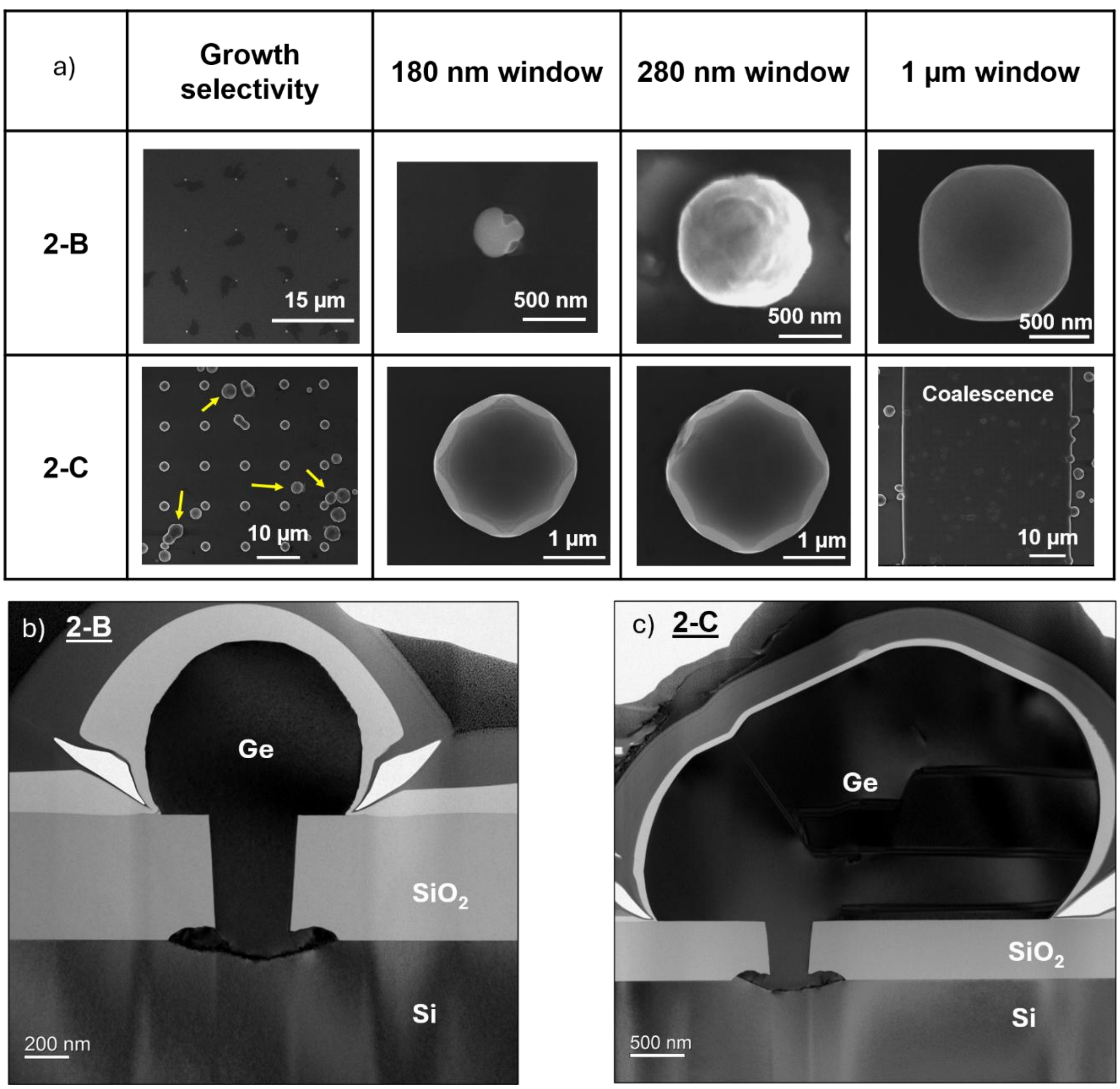

***Figure 3***. *a) Zoomed-out SEM image to assess growth selectivity, alongside SEM images of ART islands grown from 180 nm, 280 nm and 1 µm oxide windows of samples 2-B and 2-C. Yellow arrows indicated nucleation on oxide surface. Coalesced overgrowth was observed in 1 µm window of sample 2-C. b), c) TEM images on ART islands grown on 280 nm oxide window from sample 2-B and 2-C, respectively.*

Based on the reappearance of octagon, multi-facets shape in sample 2-C, further ART samples were grown with reduced HCl flux to reproduce the condition of low growth selectivity, and to assess whether GeSn growth or at least Sn droplets can occur in these cases. Results from four samples in growth round 2-D were shown in **Figure 4**: Sn droplets were indeed observed on all four samples, exclusively for ART islands grown on 1 µm oxide window. Multi-facets shapes, with circle or polygon base were observed on all samples; on sample 2-D (1), it was less sharp and can be considered as a transition from dome shape to multi-facets. Similar to sample 2-A from Ge core/ GeSn shell growth design, most Sn droplets appeared on the periphery of the ART island or on the top surface, corresponded to {100}, {110} or {311} crystalline planes.

Sample 2-D (4) showed noticeable difference compared to the other three samples: ART island shape from 180 nm and 280 nm windows evolved into a quasi-pyramid form with square base, while Sn droplets size in 1 µm window also increased, with most of them positioned on the inclined facets. With another growth attempt in sample 3-A (**Figure 5**), we were able to reproduce the pyramid form for ART island, this time with Sn droplets appearing on all three window sizes. Interestingly, the ratio of Sn droplets size over ART island size noticeably increased when they appeared in the inclined plane in sample 2-D (4) and 3-A instead of the side planes {100} and {110} like other samples, suggesting a better condition for Sn nucleation and possibly GeSn nucleation under pyramid shape compared to multi-facets shape.

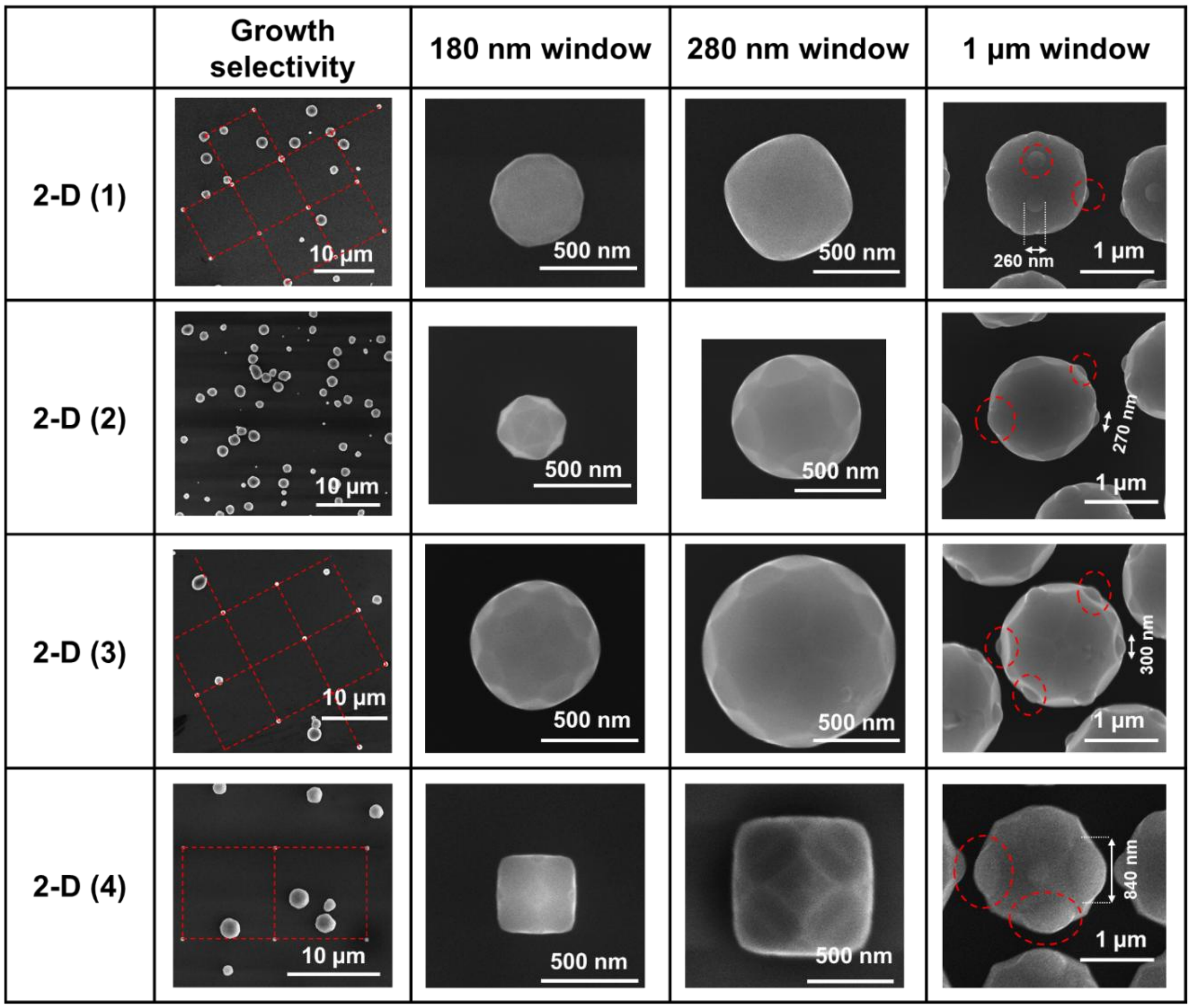

***Figure 4***. *Zoomed-out SEM image to assess growth selectivity, alongside SEM images of ART islands grown from 180 nm, 280 nm and 1 µm oxide windows of four samples in growth round 2-D. Red dashed circles helped identifying Sn droplets on ART island. White double-arrows indicated some droplet sizes, ranging from 260 nm in sample 2-D (1) to 840 nm in sample 2-D (4). Dashed lines in growth selectivity column helped identify the ART mask pattern.*

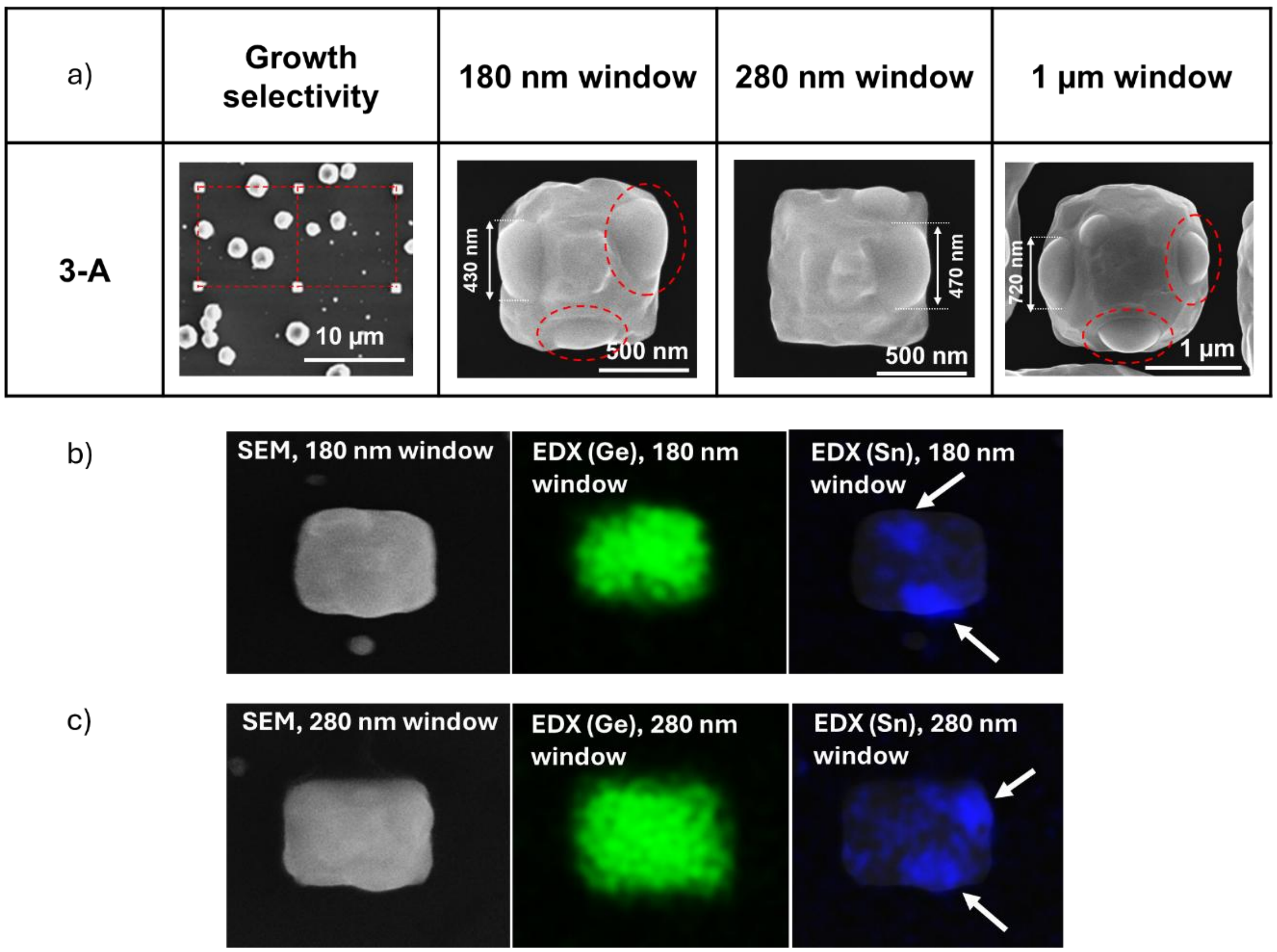

***Figure 5***. *a) Zoomed-out SEM image to assess growth selectivity, alongside SEM images of ART islands grown from 180 nm, 280 nm and 1 µm oxide windows of sample 3-A. Droplet size varied from 430 nm to 720 nm. b), c) SEM-EDX analysis on ART island grown on 180 nm and 280 nm window of sample 3-A, respectively. From left to right: Low resolution SEM image of an ART island undergoing EDX analysis, Ge (green) and Sn (blue) atomic distribution profile. White arrows pointed to Sn droplets, and transparent SEM photo overlapped with Sn EDX profile for visualization purpose. Please note that SEM-EDX analysis were conducted on different ART islands than those shown in a).*

***Summary of results from growth rounds 2 and 3***

Results from growth rounds 2 and 3 revealed two favorable conditions for Sn nucleation in ART growth: Ge core/GeSn shell, or low growth selectivity. It suggested different Sn/GeSn nucleation kinetics inside and outside of the oxide window. Sn adatoms deposition inside or near the oxide window might be more difficult, leading to a slower Sn/GeSn nucleation rate. Low growth selectivity needed to be applied in this case, to reduce the etching rate and preserve Sn/GeSn nucleation. Sn adatoms deposition might become more favorable when the grown layer was distanced from the oxide window: Sn/GeSn nucleation in this case can withstand high etching rate, which was suggested from the results of sample 2-A of Ge core/ GeSn shell growth attempt.

Still, only Sn droplets were observed in rounds 2 and 3, suggesting that optimizing GeSn ART growth recipe was more complex than simply further reducing the etching rate via HCl flux. As pointed out in Refs. [28–35] , SAG and ART growth rate increased as window size shrank, a phenomenon known as "loading effect": for GeSn SAG and ART growth, such phenomenon can limit or inhibit Sn incorporation in high Sn growth recipes, which were normally conducted under low growth rate regime limited by the growth temperature [15,36,37]. Optimized growth selectivity level for a successful GeSn ART growth should be somewhere in the middle, not too high to avoid prohibiting GeSn nucleation, but also not too low as we needed to compensate for the loading effect. Finally, results from samples 2-D (4) and 3-A revealed a correlation between better condition for Sn/GeSn nucleation and the pyramid shape, which will be further discussed in detail in the next section.

**Demonstration of GeSn ART growth (rounds 4-5)**

Using similar growth approach of short Ge nucleation combined with low growth selectivity, we obtained three more samples (4-A, 4-B and 5-A) with pyramid shape for ART islands emerging from 180 nm and 280 nm windows (**Figure 6a**). SEM-EDX results on sample 4-B and 5-A suggested a uniform Sn incorporation for ART growth on 280 nm windows. Micro-Raman spectroscopy (**Figures 6b,c**) was also conducted on all three samples, showing a red-shift of Ge($F_{2g}$) mode compared to a reference Ge ART sample, offering further verification of Sn incorporation in these samples. However, micro-Raman results on different areas on the wafer revealed a dispersion of Ge($F_{2g}$) Raman shift, suggesting that the growth conditions might not be uniform across the wafer surface, leading to a fluctuation of Sn content (**Figures 6d,e**). Random nucleation of Ge/Sn/GeSn on $SiO_2$ surface might locally impact the precursor gas flow and thus Sn incorporation in neighboring ART islands. In addition, direct estimation of Sn content from Raman results remained difficult, as Raman shift depended on both Sn content and residual strain, with the latter applied in inclined plane due to the ART island shape: so far, Raman coefficient for strained GeSn was only known for biaxial strain applied in [100] direction, corresponded to thin film GeSn grown on Ge buffer/Si substrate [38–40].

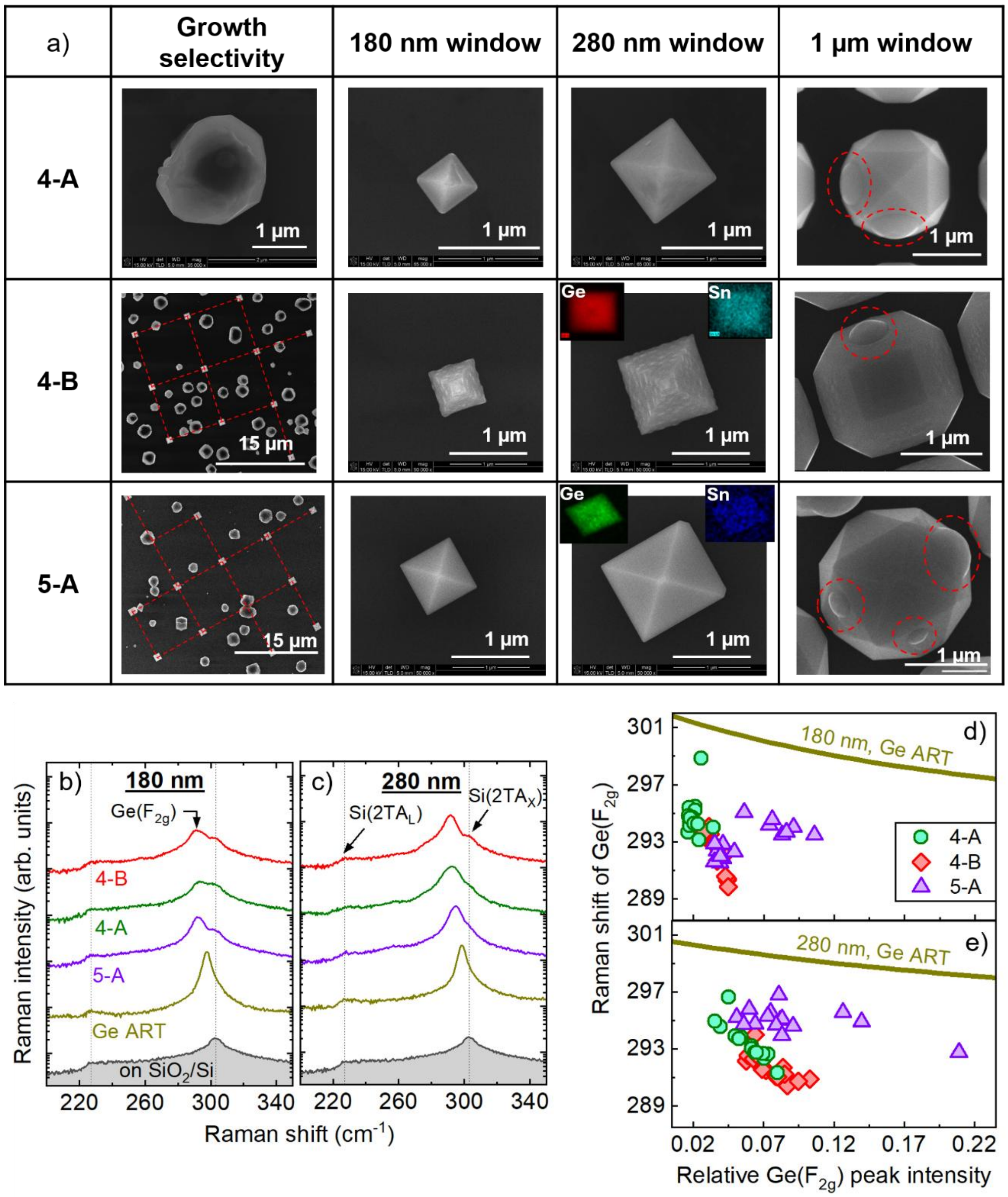

***Figure 6***. *a) Zoomed-out SEM image to assess growth selectivity, alongside SEM images of ART islands grown from 180 nm, 280 nm and 1 µm oxide windows of samples 4-A, 4-B and 5-A. EDX was conducted on ART islands grown from 280 nm window in samples 4-B and 5-A, showing a uniform Sn distribution. b), c) Micro-Raman spectra for ART islands in samples 4-A, 4-B, 5-A,*

*grown in 180 nm and 280 nm windows respectively. Results were compared with micro-Raman spectra of Ge ART reference sample and those of $SiO_2$ template on Si substrate. d), e) Plots of Raman shift as function of relative Ge($F_{2g}$) peak intensity for a variety of ART islands grown on 180 nm and 280 nm windows respectively, compared to Ge ART results. The relative Ge($F_{2g}$) peak intensity was calculated as the ratio between the intensities of $F_{2g}$ Raman modes of Ge and Si (observed at 520 $cm^{-1}$, not shown in the Raman spectra).*

To gain further insight into GeSn layer structure, we conducted TEM and EDX analysis on samples 4-A, 4-B and 5-A, with results shown in **Figure 7**. A 54.6° angle was measured between the ART island facets and (100) plane, identifying them as {111} facets. For samples 4-A and 4-B (**Figures 7a,b**), TEM-EDX analysis revealed a Ge core/ GeSn shell structure on both samples, instead of the nominal design of bulk GeSn with short Ge nucleation: this self-induced core/shell structure was a stable configuration for GeSn ART growth, aligned with previous observation in sample 2-A and resembling some previous works on InGaAs and InGaN nanostructure growth [41–48]. Zoom on the Ge/Si interface revealed an array of misfit dislocations with even spacing of 10 nm, fully consistent with the theoretical spacing value $d = \frac{b_{Ge}}{\varepsilon}$, where $b_{Ge} = 4.0$ Å was the Burger's vector of Ge, and $\varepsilon = 4\%$ was the misfit strain due to lattice mismatch between Ge and Si, clearly showing that the Ge layer was fully relaxed. Similar to Ge ART reference sample 1-A, noticeable under-etching into the Si substrate was observed at the base of the oxide window, with threading dislocation trapped by the oxide walls instead of propagating into the upper grown layer. High-resolution TEM (HR-TEM) revealed different layers on the GeSn shell on sample 4-B, suggesting a change of Sn content as GeSn layer built up. For sample 5-A (**Figures 7c**), TEM-EDX analysis revealed bulk Sn distribution in the ART island, with weaker intensity compared to GeSn shell layer in samples 4-A and 4-B, however.

Cut section of TEM-EDX maps (**Figure 8**) quantified the Sn content distribution in GeSn ART samples. On Ge core/ GeSn shell ART structure, Sn content increased up to 6% and 8% in samples 4-A and 4-B, respectively. On bulk GeSn ART structure (sample 5-A), TEM-EDX data showed weaker Sn content below 1% in the bulk, before sharply increasing on the pyramid edge. It should be noted that TEM-EDX results were limited to individual ART island and might not be representative for the full ART wafer in each case, as suggested by the scattered micro-Raman data observed earlier: it was possible that bulk GeSn and Ge core/GeSn shell ART islands co-existed on a same wafer, under local fluctuation of growth condition.

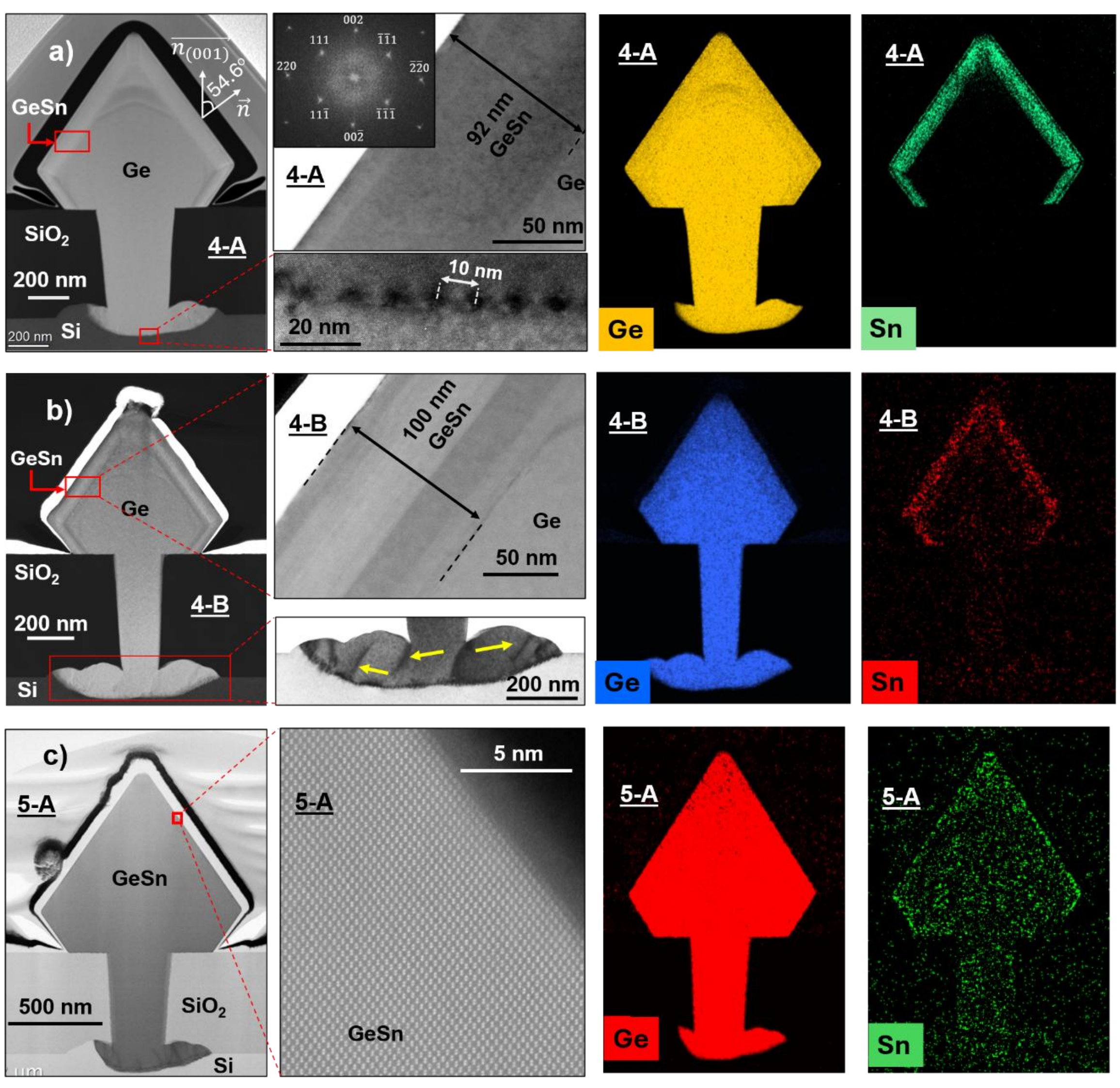

***Figure** 7. TEM and EDX analysis of a) sample 4-A, 280 nm window, b) sample 4-B, 180 nm window and c) sample 5-A, 280 nm window. Angle between (111) facet and (001) plane, alongside diffraction map and spacing between misfit dislocation at Ge/Si interface was provided in TEM images of sample 4-A. Zoom showing threading dislocation distribution and on underetched area (identified with yellow arrows) was provided in TEM image of sample 4-B. HR-TEM image on atomic distribution in bulk GeSn was provided for sample 5-A.*

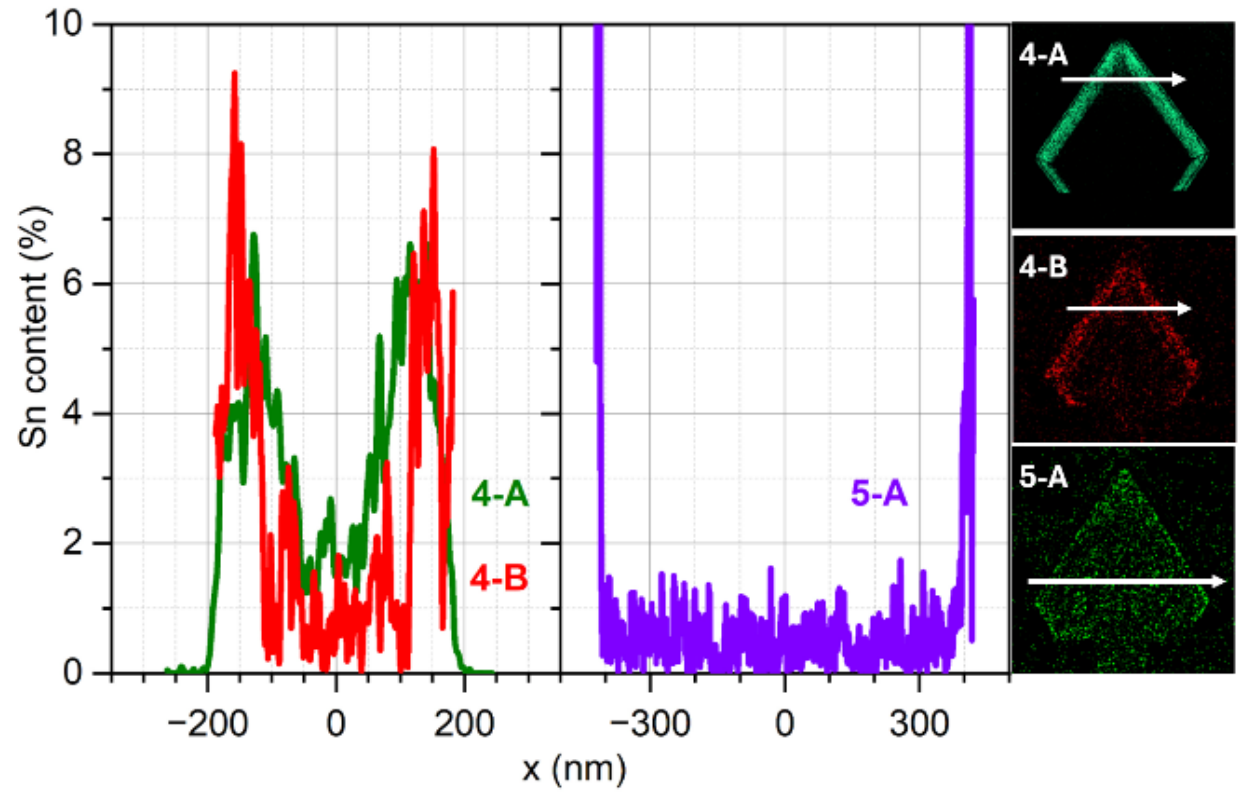

***Figure 8**. Sn content extracted from TEM-EDX cut section for samples 4-A, 4-B, and 5-A. Cut sections were indicated with white arrows on insets.*

Results from rounds 4 and 5 verified the feasibility of GeSn ART growth with medium to high Sn content and offered a detailed look into its growth kinetics. First, pyramid shape, with {111} facets of lowest surface energy for diamond/face-centered cubic lattice [49] was confirmed to be the preferable configuration for GeSn growth. Based on previous work on Stranski-Krastanov growth of Ge nanocrystals on Si [50], multi-facets/dome and pyramid shapes corresponded to two minima in surface energy landscape, thus two stable configurations for either nanocrystal or ART island. In GeSn ART growth, this pyramid shape might be explained by an increase of growth rate ratio between <111> directions and other crystalline direction compared to Ge ART growth.

Secondly, with our current growth recipe, medium to high Sn content (6% - 8%) existed in shell form, while bulk GeSn configuration can only be confirmed below 1% Sn. This bulk Sn content was close to the maximum solubility of Sn in GeSn alloy of 1.12 % Sn under thermodynamic equilibrium [51]. Besides the adjustment of growth rate through growth selectivity described in previous section, local growth temperature can also be a deciding factor for high Sn incorporation in GeSn ART island. Here, $SiO_2$ mask of lower thermal conductivity (~ 1.5 $W.m^{-1}.K^{-1}$) than Ge (~ 60 $W.m^{-1}.K^{-1}$), combined with restricted air flow can locally trap heat and increase the temperature inside narrow, deep oxide window during ART growth. This local heating can inhibit high Sn incorporation in the bulk, as it was highly sensitive to temperature fluctuation: in thin film growth, Sn content can reduce from 17% Sn to 0% Sn with modest increase of growth temperature from 300 ℃ to 350 ℃. This temperature range was consistent with previous estimation of overheating up to 45 ℃ in InAs quantum dots growth on Si substrate with oxide template [52]. Once overgrowth outside of $SiO_2$ took place, local heating effect might become less pronounced due to unobstructed air flow and thus improved heat exchange through convection; high Sn GeSn can thus start nucleating and formed shell layer as observed in samples 4-A and 4-B. Lowering the growth temperature can be an efficient pathway to obtain high Sn incorporation in bulk GeSn ART, and can also lead to different GeSn ART structure: gradient increase of Sn content with constant growth temperature due to different local heating effect inside and outside of $SiO_2$ windows, or constant Sn content with different growth temperatures applied during confined growth phase and overgrowth phase, to compensate for the change of local heating effect.

## CONCLUSION

In this paper, we reported GeSn ART growth with two observed configurations: self-induced Ge core/ GeSn shell at medium Sn content of 6% - 8% Sn, and bulk GeSn at low Sn content below 1% Sn. Pyramid form with {111} facets was found to be a reliable indicator of successful GeSn growth, and low growth selectivity was found to increase the chance of GeSn ART growth success. In addition to the growth selectivity tuning, future work will also focus on adjusting the growth temperature to further understand the impact of local heating on bulk Sn incorporation in GeSn ART growth, targeting bulk GeSn ART island with high Sn content.

## METHODS

### Material growth and characterization

Standard lithography was used to imprint ART window pattern on $SiO_2$ layer. The $SiO_2$ layer was formed using a thermal oxidation process at 1000 ℃. Ge and GeSn ART samples was then grown using a ASM Epsilon® 2000 RPCVD reactor, with $GeH_4$ and $SnCl_4$ as Ge and Sn gas precursors. SEM characterization was performed using FEI Nova Nanolab 200 system, with 15 kV of accelerating voltage. SEM-EDX analysis was performed using FEI-Nova NanoLab 200 system and Emcrafts Cube II system. STEM-EDX characterization was performed using Hitachi HD2700 AC-STEM system with 200 kV of accelerating voltage. Micro-Raman measurements were conducted using a Horiba Jobin-Yvon LabRam spectrometer with 632.8 nm laser excitation, equipped with microscope systems and a 100× objective lens, resulting in 2 μm diameter focused laser spot.

## AUTHOR INFORMATION

**Corresponding Author**

* Shui-Qing Yu

Email: syu@uark.edu

**Author Contributions**

S.-Q. Y., W. D., G. S, B. L. and J. L. proposed and supervised the projects. M. C. and S. M. produced SiO2 patterned wafers for ART growth. H. S. and P. C. G conducted SEM-EDX measurements. F. M. O. conducted micro-Raman measurements. M. B. performed TEM measurement. H. S. and Q. M. T. conducted data analysis and wrote the manuscript. All authors discussed the results and commented on the manuscript.

‡ H. S and Q. M. T contributed equally to this work

**Notes**

The authors declare no competing interests.

## ACKNOWLEDGMENT

This work is supported by United States Air Force under Contract No. FA865023C1140, the Department of Defense Small Business Technology Transfer (STTR) Program, Office of the Secretary of Defense-Basic Research Office (OSD-BRO) and the Army Research Office under Contract No. W911NF-24-C-0052, the Air Force Research Laboratory (AFRL/RYDHS and AFRL/RYKSE) under Contract No. FA237724CB033, and Office of Naval Research under Grant No. N000142412651.

Lincoln Laboratory at Massachusetts Institute of Technology would like to acknowledge the support as follows: DISTRIBUTION STATEMENT A. Approved for public release. Distribution is unlimited. This material is based upon work supported by the Department of the Air Force under Air Force Contract No. FA8702-15-D-0001 or FA8702-25-D-B002. Any opinions, findings, conclusions or recommendations expressed in this material are those of the author(s) and do not necessarily reflect the views of the Department of the Air Force. 

We would also like to acknowledge Grey Abernathy for his support on SEM-EDX and TEM measurements.